**Titles**

- Full title: Wormholes grow along paths with minimal cumulative surface
- Short title: Path selection during wormhole growth

**Authors**

Y. Yang,[1]* S.S. Hakim,[1] S. Bruns,[1] K. Uesugi,[2] K. N. Dalby,[1] S.L.S. Stipp,[1] H.O. Sørensen[1]

**Affiliations**

[1] Nano-Science Center, Department of Chemistry, University of Copenhagen, Universitetsparken 5, DK-2100 Copenhagen, Denmark.

[2] SPring-8, Mikazuki, Hyogo 679-5198, Japan

* Corresponding author: yiyang@nano.ku.dk

**Abstract**

Growth of wormholes in porous media can lead to self-organization of flow networks with an overwhelming geometric complexity. Despite decades of study, the mechanism by which a dominant wormhole develops its path during growth remains elusive. Here we show that the trajectory of a growing wormhole can be predicted by identifying the flowpath with a so-called minimum cumulative surface. Our theoretical analysis indicates that the cumulative surface determines the position of the dissolution front. We then show, using numerical simulation based on greyscale nanotomography data, that the tip of an advancing pore always follows the migration of the most far reaching dissolution front. Finally, we show the good accord between theory and the observation with *in situ* microtomography. Our result suggests that wormholing is a deterministic process rather than, as classical theories imply, a stochastic one that can only be analysed statistically. Our results shed new light on engineering of artificial flow systems by exploiting self-organization in natural porous materials.

## Introduction

The growth of wormholes in porous media, or wormholing, is a ubiquitous self-organization phenomenon in earth sciences (*1-6*). It results from a positive feedback between fluid flow and mineral dissolution (*7-10*). When a reactive fluid flows through a dissolving porous medium, it prefers to erode the regions it has previously etched (*11*). This process, called dissolutive percolation, produces highly conductive channels (wormholes) and leads to a stratified distribution of fluids (Fig. 1) (*12*).

Understanding the mechanism of wormhole growth is of fundamental importance in numerous scientific disciplines. For example, in geochemistry, all kinetic measurements of water-rock interactions in porous media suffer from a classic surface area-associated normalization problem (*13-16*). The development of bypassing channels can render a significant portion of the material "unreactive" – a result of insufficient solid-liquid contact. This portion therefore contributes to the erroneous estimate of surface area and chemical affinity used in rate normalization (*17*). In geologic carbon sequestration and oil field stimulation, the injection of acidic fluids generates highly complex flow networks (*18-24*). Managing the development of such networks, i.e., controlling the rate of network expansion while regulating the size distribution of the growing channels, is essential for appropriately dissipating $CO_2$ into aquifers (*13, 21, 25-29*) and for effectively recovering hydrocarbons from reservoirs (*12, 30, 31*). Additional examples include nuclear waste management and environmental contaminant immobilization, where an uncontained development of advective paths in soil may lead to disastrous consequences (*32, 33*).

Despite decades of fruitful research, the exact mechanism of wormhole growth remains elusive. The problem is best formulated in the context of reactive infiltration instability (RII), where a migrating dissolution front is morphologically unstable to perturbations in transport properties (*1, 3, 34-37*). Regional heterogeneities in porous media are one such typical perturbation. The instability is determined by the strength of the one-way coupling of three spatially distributed parameters: velocity, permeability and reactant concentrations. The spatial distribution of fluid velocity (flow field) depends on the local permeability (microstructure), while the latter further depends on the distribution of reactant (reactivity field) through chemical kinetics. To close this loop of positive feedback, the reactivity field depends on the flow field as a result of reactant transport. If any of these dependencies vanishes or has the opposite effect, the feedback loop breaks and wormholes are not formed. For example, if there is no imposed flow (or the flow rate is negligible), the dissolution front remains planar and moves as a plug flow, a phenomenon sometimes termed "face dissolution" (*30, 38-41*). Similarly, a zeroth order dissolution reaction, even if very fast, does not generate wormholes because its rate is independent of the reactivity field.

The essence of the positive feedback is the amplification of heterogeneities in porous media. Therefore, any negative feedback that reduces differences in reactive transport properties hinders the formation of wormholes. Noticeable feedbacks of such include solid availability, mixing, mineral precipitation and fluid compressibility. Solid availability imposes an upper limit on regional porosity. If two points in space have different dissolution rates, they will only end up with different porosities if they do not hit this upper limit (i.e. porosity = 1) simultaneously. As a result, "face dissolution" can also be observed when the acid capacity (the ratio of mineral solubility to its molar density) is large. This type of face dissolution is fundamentally different from one that stems from the absence of an imposed flow field. The former has a weakly unstable dissolution front (*11, 42, 43*) while the latter is a stable system with a strictly planar dissolution front in numerical simulation (*44, 45*). Mixing reduces spatial reactivity differences. The most important mixing mechanism is molecular diffusion (*46, 47*). In oil field stimulation, highly viscous acids are used to inhibit molecular mixing induced radial dispersion in order to enhance wormhole growth (*48*). The significant impact of diffusion has also motivated the use of the Péclet number (Pe), the ratio between advective to diffusive mass transfers, as one of the decisive variables in describing wormholing patterns (*40, 41, 43, 45, 49*). Occasionally, the product of Péclet number and Damköhler number (Da, the ratio between residence time and characteristic reaction time) is used to quantify the competition between radial dispersion and dissolution (*50-52*). The impact of mechanical dispersion has also been investigated numerically (*53*). Mineral precipitation has the opposite effect in changing regional porosity compared to dissolution (*31, 46, 51, 54, 55*). This negative feedback has motivated the use of *in situ* cross-linked acids to stabilize the dissolution front and prevent unnecessary fluid channeling in matrix acidization (*56*). Lastly, if a percolating fluid is compressible, it serves as a buffer that smooths out the pressure gradient. Consequently, a greater fluid compressibility leads to a stabilized dissolution front (*36*). It is worth emphasizing that inhibiting wormhole morphologies is different from inhibiting dissolution. All the aforementioned mechanisms inhibit wormhole formation by reducing the strength of the positive couplings. This is qualitatively reflected in the size and shape of wormholes (shorter and thicker). However, for example, mixing can enhance mineral dissolution at the same time, if the dependence of the reaction rate on reactant concentration is convex (typical for mineral dissolution) (*57-61*).

A central question remains: is wormholing deterministic? Can we predict the growth of a wormhole, or is the process stochastic and only described statistically? If it is a random process, as is suggested by many pioneers in the field (*42*), then classifying wormholing patterns qualitatively with the help of dimensionless numbers (e.g., Da, Pe, Da·Pe and the Zhao number) constitutes the optimum strategy. If, however, the deterministic nature of the problem is only hindered by mathematical difficulties and the challenges in characterizing the microstructures of porous media (*62*), then the numerical and imaging techniques available today present an opportunity to thoroughly reformulate the problem (*26, 63, 64*). Here we combine nano-computerised tomography (nanoCT), grey scale numerical simulation, and *in situ* microtomography (*in situ* μCT) with a theoretical analysis to show that the path of the dominant wormhole is pre-deterministic in a system subject to reactive infiltration instability. The prediction relies on identifying the flow path with the most far-reaching dissolution front. Our result constitutes the first step of describing the self-organization of flow system in natural materials as a population balancing problem: a problem of quantifying the dynamics of the number of channels. This number is initially determined by the abundance of heterogeneities but ultimately limited by solid availability as the microstructure evolves.

## Results

### Pore growth follows the migration of the reaction front

We simulated the microstructural evolution of a chalk cuboid ($5\times5\times25$ μm$^3$) during dissolutive percolation (Fig. 2). The simulation is based on a digital model of chalk microstructure obtained by X-ray holotomography with voxels having a 25 nm edge length. We chose to use nanotomography (nanoCT) data in the numerical simulations because of its detailed characterisation of the geometric surface within the sample. The 3D matrix of voxels is treated as a reactor network with imposed fluid flow. The permeability and tortuosity of each voxel are expanded as functions of its voidspace and truncated beyond the second order term, leading to a parabolic dependence of a phenomenological coefficient on voxel porosity. This coefficient is then calibrated with experimentally measured permeability values of chalk and used with Darcy's law as the governing equation for the pressure drop across individual voxels. This approach allowed us to implement the 32-bit grey scale data directly in numerical simulations and avoid the significant information loss during image segmentation (*65*). The flow field is imposed by assigning the injection of fluid as source terms in the inlet reactors and the effluent removal as sink terms in the outlet reactors (Fig. 2A). The equivalent free-stream fluid velocity was 5 $\mu$m/s. The injection fluid is pre-equilibrated with $CO_2$ (1 bar, 25 $^o$C). Degassing of $CO_2$ during percolation is neglected therefore the fluid viscosity does not vary in space. The rate law of calcite dissolution was used to approximate the kinetics of chalk dissolution.

The growth of pore structures was sensitive to regional variations in porosity (Fig. 2B). The tip of the wormhole showed a pronounced tendency to evade less permeable regions as it advanced towards the outlet. This phenomenon is characteristic of the development of flow systems induced by reactive infiltration instability (RII), where the migration of the reaction front deviates significantly from a unidirectional plug flow. The structural evolution stemming from mineral dissolution provides a positive feedback between the rates of flow and chemical reactions. This positive feedback ensures that the flow path with the most far reaching reaction front increases its reactivity compared with other flow paths over time. This "Matthew Effect" amplifies regional heterogeneities in a porous matrix, produces solid channelisation and fluid focusing, and results in the breakthrough of a porous medium without dissolving the entirety of its solid material. In the simulation the breakthrough took place between porosity 0.4 and 0.6 (Fig. 2C) and was manifested by a sharp decrease in the pressure drop along the flow direction.

The evolution of geometric (GSA) and reactive surface area (RSA) also demonstrates the characteristics of an RII-dominant process (Fig. 2C). The geometric surface reflects the amplitude and frequency of the spatial variations of material density. This variation represents a regional heterogeneity that can be amplified by RII. The GSA therefore increases until the occurrence of a breakthrough. After breakthrough the dissolution pattern is dominated by the expansion of the major flow channel and the GSA decreases as solid material depletes regionally. This non-monotonic development of geometric surface area thus reflects a dynamic balance between the advancing of the pore tip in the flow direction and the expansion of pore volume perpendicular to the direction of flow. The former increases regional density contrast and thus the GSA, while the latter is limited by the locally available amount of solid material and decreases the GSA. The reactive surface is the portion of the geometric surface that serves as the interface for water-rock interactions. RSA is separated from the unreactive surface by the reaction front. In a dissolutive percolation RSA depends heavily on the dispersion of reactive fluid near the tips of the advancing pores. The slower the fluid reactivity depletes along a streamline, the farther the reaction front extends and the greater the reactive subvolume. Consequently, the apparent solubility of the solid (the total amount of mineral dissolvable in a closed free-drifting system), the fluid residence time and the geometric surface area available along a flow path determine simultaneously the reactive surface. RSA ceases to increase as the reaction front extends beyond the fluid outlet and becomes equal to the area of the isosurface for porosity = 0. This isosurface separates the wormhole from the rest of the porous medium (Fig. 3A).

The simulation indicates the growth of wormholes follows the migration of the reaction front for two reasons. First, the reaction front encompasses the reactive subvolume of the medium. At any given instant wormholes can only grow within this confined space. Second, the flow path with the farthest reaching reaction front also dissolves fastest (characteristic of RII). Two important implications follow: 1) both the reaction front migration and the pore growth follow streamlines in a flow field. On the same streamline, the tip of a pore has the highest reactivity while the reaction front has the lowest reactivity. 2) The streamline with the largest initial distance between the pore tip and the reaction front (the "Damköhler length") is the axis of wormhole growth. This is because of the spontaneous fluid focusing during microstructural evolution.

**The cumulative surface determines the position of the reaction front**

The position of the reaction front can be computed by considering streamlines as flow paths with infinitesimal cross sections and by defining cumulative surface (CS) as

$$\text{Cumulative Surface} = \int_s SSA \cdot ds / \mathbf{v}, \quad (1)$$

where $s$ indicates an integral along a streamline, $SSA$ (m$^{-1}$) represents the position dependent specific surface area and $\mathbf{v}$ the velocity. $ds/v$ gives $d\tau$ which is the residence time of a fluid element travelling along the integrated path (s). A probability density function of CS for the complete flow field can then be evaluated by applying Eq. 1 to a sufficiently large number of streamlines that originate from the fluid inlet. In the inset of Fig. 3A we show the shape of a probability distribution based on the CS of 40,000 streamlines (one from each inlet voxel) at the beginning of the percolation. Given the extent of mixing, the total amount of dissolved calcite can be calculated directly from this distribution. For convenience and without losing generality, we assume segregated flow in the following analysis. This assumption means fluids flowing in different streamlines do not exchange mass or energy. Consequently, it follows for each streamline

$$\int_0^\tau SSA \cdot dt = \int_0^{[Ca^{2+}]} \frac{d[Ca^{2+}]}{r_{diss}(\text{pH}, \text{SI})}, \quad (2)$$

where $\int_0^\tau SSA \cdot dt$ is the cumulative surface for a residence time of $\tau$ (s/m), $[Ca^{2+}]$ represents the aqueous concentration of calcium cations (M) and $r_{diss}$ is the rate of calcite dissolution (mol·m$^{-2}$·s$^{-1}$) that depends on the SI (saturation index) of calcite and the pH. Releasing calcium from the solid increases pH and the SI, which decrease the dissolution rate. According to the RHS of Eq. 2, the reaction front extents farther along a flow path with a lower chemical conversion. Combined with Eq. 1 and the nature of RII, we conclude that wormhole growth favors streamlines with a minimal cumulative surface. Such streamlines have low chemical conversion rates and a far reaching reaction front. More importantly, both features are preserved and amplified by RII as the pore structure evolves.

The impact of the cumulative surface on the pore development is shown in Fig. 3. The morphology of the wormhole at the cusp of breakthrough (porosity = 0.44) resembles closely the shape of the streamlines drawn based on the initial flow field (Fig. 3A). These streamlines are in the lower decile of cumulative surface (inset red bars) at the beginning of the percolation. This resemblance suggests the possibility of predicting the potential trajectory of wormhole growth in any given microstructure. In Fig. 3B we show 5 streamlines in the initial flow field when the overall porosity = 0.20. For each we computed the CS using Eq. 1 and the dissolution rate using Eq. 2 (Fig. 3C). The latter reveals the positions of the reaction fronts. For example, the grey dashed line cuts the curves at the required travel distance for decreasing the dissolution rate to 5 $\mu mol \cdot m^{-2} \cdot s^{-1}$. If this (arbitrary) value represents that of the reaction front, the yellow streamline would have the shortest reactive length (~10 μm) while the red one has the longest (~27 μm). Note that with the initial microstructure no wormhole exists and therefore its tip is the fluid inlet (at $s = 0$ μm). The red streamline is expected to become the path of wormhole growth because it has the largest distance Damköhler length. If this sieving process is repeated for streamlines representative of the complete initial flow field, the axis of the growing wormhole can be identified without additional numerical simulations. The caveat is that the relocation of grains with the stream or the mechanical sustainability of the structure is ignored. In this study, we also did not consider the reactivity difference between the biogenic materials (coccospheres) and the abiotic calcite. Although they are both calcium carbonates, the presence of organics on the former surface, e.g. polysaccharides, can modify its dissolution kinetics.

**Preferential dissolution recorded by *in situ* μCT linked to lower cumulative surface**

We recorded the development of chalk pore structure during dissolution using *in situ* X-ray microtomography (μCT). The highest achievable voxel size with the *in situ* setup was 248 nm. Although with this resolution the geometric surface area could not be as accurately characterized as with nanoCT, the *in situ* scans allowed us to track the evolution of the pores and distinguish the preexisting ones from those developed during dissolution. In Fig. 4 we show 200 × 245 $\mu m^2$ cross sections of a time series collected over a 21 hour percolation experiment. The direction of fluid flow is normal to the image plane and pointing inwards. Perspective 3D views of the microstructure (200 × 245 × 300 $\mu m^3$) before and after the percolation are given in Figs. 5A and 5B. The major flow channel advanced gradually in the flow direction and expanded in diameter from the lower right of the region of interest (ROI). The preferential removal of material is demonstrated in the distinct dissolution patterns within the ROI. Near the channel wall are flow paths with shorter residence time and thus lower cumulative surface. Fluid in this region is farther from equilibrium and therefore dissolves chalk faster. In contrast, the streamlines going through the porous area on the left side of the ROI are expected to have a higher cumulative surface. The fluid following these paths was therefore closer to saturation and dissolved chalk slowly. A comparison between Figs. 4A and 4J shows that the left side of the ROI did become more porous with time, but the fine grainy texture had been preserved (see also Fig. 5B). As time elapsed more fluid was redirected to the right (because of the greater permeability). This flow stratification further increased the residence time in the less porous region and resulted in the positive feedback characteristic to RII. Overall, the *in situ* observation is consistent with the qualitative behaviors observed for the numerically simulated system. The evolution of the microstructure met the expectation of a cumulative surface-based analysis, although the size of the developing flow channel was significantly greater than the simulated wormhole because of a greater flowrate (0.01 ml/min).

The aforementioned screening procedure (Fig. 3) was applied to the microstructure reconstructed from the μCT datasets to identify the potential pathways of pore growth. In Fig. 5C we show the streamlines with the lowest 10% of cumulative surface at the beginning of the experiment. The lowest 2% are colored red and are guiding the wormhole development. This serves as a rough estimate for two reasons. Firstly μCT cannot fully resolve the spatial variations of density in fine grain materials such as chalk. The cumulative surface is therefore underestimated. Secondly the ROI only covers a subvolume of the cylindrical sample. The numerically imposed flow field thus does not fully reproduce the distribution of streamlines in the actual experiment. Nevertheless, the clustering of the streamlines in the intact microstructure strongly suggests wormhole formation near the lower right of the ROI. In Fig. 5D we draw streamlines originating from the same points as in 5C but based on an updated flow field. This flow field is computed by imposing a pressure difference across the microstructure after the percolation. Red color indicates streamlines with the lowest cumulative surface. Fluid in these streamlines bypasses the porous material completely and does not contribute to structural dissolution. The effect of CS on path selection is manifested clearly in the change of the probability distribution, shown in Figs. 5E and 5F. The original heterogeneity of chalk produces a distribution of CS within 2 orders of magnitude, with a mean of $1.23 \times 10^7$ s/m and a standard deviation of $8.43 \times 10^6$ s/m. After wormhole formation the probability density function shows a typical bimodal distribution, with one mode near zero representing the bypassing flow and the other mode at a large CS. The probability density function in Fig. 5F covers 14 orders of magnitude and yields an average CS of $4.0 \times 10^{10}$ s/m and a standard deviation of $7.25 \times 10^{11}$ s/m.

## Discussion

Wormhole growth favors flow paths with minimal cumulative surface. This preference in path selection is a result of the concerted functioning of three independent mechanisms. Firstly, the tip of a growing pore follows the migration of the reaction front on the same streamline as a result of solid material depletion. Secondly, the distance between the pore tip and the reaction front is largest when the cumulative surface along the streamline is lowest. This is because of the interplay between reaction rate, residence time and surface area. Lastly the positive feedback between flow and mineral dissolution, determined by infiltration instability, amplifies the difference between the positions of reaction front over time. Consequently the streamline with the minimal cumulative surface in the initial microstructure is the axis of wormhole growth.

The implications are twofold. The finding presents new challenges in predicting the dynamics of fluids in ever evolving porous media. The morphing of microstructures and the amplification of heterogeneities in natural materials are extremely challenging to incorporate into reactive transport modeling. If the migration of, e.g., supercritical $CO_2$, contaminants or remediation agents are accompanied by a modification of pore geometry, new numerical schemes are needed to implement both first principles (e.g., Navier-Stokes equations) and phenomenological laws (e.g., Darcy's law) coherently in the same simulation domain. On the other hand, the unique feature of infiltration instability emphasizes the importance of topological characterisation. For example, in the presented cases if the flow path with the minimum cumulative surface can be identified in the initial geometries, neither percolation experiment nor numerical simulation with morphing boundary conditions would be required to predict the preferential dissolution pathway. The advancement of X-ray imaging techniques is therefore not only improving geometrical characterization but also allows predicting the system evolution directly.

A pronounced limitation of this study is that mechanical impact was neglected. Both the numerical simulation and the percolation experiment were conducted under conditions where the exerted stress does not lead to a collapse of the microstructure. In reality infiltration instability can be triggered by processes other than chemical reactions, such as in hydraulic fracturing. The confining pressure of a percolation has also been shown to affect microporosity. The validity of the conclusions is thus limited to scenarios where the morphologies of wormholes are not affected by the materials' mechanical strength. We also did not consider processes that may increase regional porosities in rock, e.g. the relocation of grains/particles by fluid flow, or mineral precipitation. The latter often stems from upstream rock dissolution and becomes more prominent towards downstream as cumulative surface increases. In contrast to a dissolutive percolation, precipitation in a pressure driven flow field has negative feedback between chemical reaction rate and flow velocity. This difference further poses the question whether or when a dissolution-precipitation scenario would demonstrate the unique features of reactive infiltration instability. Overall, the convolution of processes on pore scale where the first principles meet the phenomenological laws is still poorly understood, even without considering chemical heterogeneities (typically presented in sandstones).

## Materials and Methods

### X-ray computerized tomography and experimental design

**NanoCT**. Samples of chalk from the North Sea Basin (#16) of diameters ~500 μm were collected from drill cuttings or core material. The pore structure was mapped by X-ray holotomography, using the nanotomography setup at the ID22 beamline (29.49 keV) at the European Synchrotron Research Facility (ESRF) in Grenoble, France. Samples were imaged dry and in air. Volume images of 25 nm voxel were reconstructed from 1999 radiographs (360º rotation , 0.5 s exposure) using the holotomography reconstruction method (*66*). For further details see Müter et al (*67*).

**Percolation experiments and *in situ* μCT**. Aalborg chalk outcrop samples were machined into cylindrical cores of 600 μm diameter and cut to 2 mm length using a razor blade. Each cylinder is wrapped in a heat shrinking tube that was fixed, through heating and application of epoxy glue, to two pieces of stainless steel needles at each end. Both needles were 3 cm long and 500 μm in diameter. The samples were percolated in a miniature Hassler core cell. The cell has 2 pairs of inlets and outlets for confining and percolating solutions. After mounting the sample vertically inside an aluminum tube (2.5 mm OD), the cell was confined to a pressure of 10 bar. The confining side was then sealed and the percolating inlet and outlet were opened to allow the fluid to be driven through. We used MilliQ water equilibrated with 1 bar $CO_2$ under ambient temperature (25 $^oC$), leading to an injection at a pH of 3.91. The $CO_2$ gas was bubbled constantly into the reservoir of MilliQ (500 ml) throughout the experiment. A SSI series II HPLC pump (Scientific Systems, Inc.) was used to control the percolation flow rate. Time resolved microtomography experiments were conducted at the BL20XU medical beamline at SPring-8, Japan (*68*). The dissolution process was monitored by recording 21 tomographic datasets of 1800 projections exposed for 1 sec over 180° at 28 keV, resulting in a time series with a 40 min time resolution. The projections of 2048 by 2048 pixels were acquired with a 20× objective lens yielding an effective pixel size of 248 nm.

**Qualitative CT studies.** A number of qualitative studies on the wormholing phenomena, both *ex situ* and *in situ,* were conducted at various synchrotron beamlines as well as on a benchtop CT. Selected morphologies are presented in Figure 1. Figs. 1A and 1B were collected using 28 keV radiation at the P05 beamline of Petra III at Deutsches Elektronen-Synchrotron (DESY), Hamburg, Germany. 1200 projections were collected over a sample rotation of 180°, exposing each projection for 1050 ms. The radiographs were recorded with a 20× zoom lens on a CCD detector of 3056×3056 pixels (pixel size = 1.3 μm after 2-times binning). Fig. 1C was collected on an XRadia 410 benchtop CT at 100 keV. 1601 projections were collected over a sample rotation of 360°, exposing each projection for 1200 ms. The radiographs were recorded with a 10× zoom lens on a CCD detector of 1024×1024 pixels (pixel size = 2.72 μm). Figs. 1D and 1E were collected using 21.5 keV (2% bandwidth) radiation at the TOMCAT beamline at the Swiss Light Source. 1501 projections were collected over a sample rotation of 180°, exposing each projection for 250 ms. The radiographs were recorded with a 10× zoom lens on a CCD detector of 2560×2160 pixels (pixel size = 0.65 μm). The field of view was 1.66 by 1.40 mm. The image reconstruction was performed with the GridRec algorithm (*69*) using a Shepp-Logan filter (*70*), zero-padding of 0.5 and ring artefact removal on a grid with the given voxel size.

**Grey scale modeling and simulation**

The 3D reconstruction of each tomographic dataset was performed by standard filtered back-projection with the respective in-house reconstruction software. A uniform greyscale distribution in the reconstructed image stacks was assured by mapping 99.98% of the dynamic range in every reconstructed 2D slice to a 0 to 1 interval in a 32bit greyscale image. All reconstructions were then cropped because the flow cell incorporates four stabilizing stainless steel rods that shade some of the projections and thereby cause streaking artefacts at the edges of the reconstruction. The cropped reconstructions were further processed by manually estimating the noise level from homogeneous regions of void and material phases and applying four iterations of 3D iterative non local means denoising (*65, 71*). For the *in situ* μCT datasets, alignment of the 4D timeseries was performed after denoising with in-house digital volume correlation software using Pearson's correlation coefficient as a quality metric.

Degassing of $CO_2$ causes the fluid density to vary from inlet to outlet, i.e. local voxel porosity can only be approximated. This was done by fitting a Gaussian mixture model to the greyscale histogram and assigning a soft segmentation into pore space, high porosity phase, low porosity phase and nonporous material phase. Expected mean greyvalues for the void and material phase were determined by tracking voxels that do not change intensity over time. The expected porosity for the mix phases was then determined by linear interpolation between void and material phase. For the microtomography reconstructions in Fig. 5 the fuzzy phase assignment was further refined by optimizing a Markov random field model where the interface potentials between the four phases were defined by their difference in expected material content. Ultimately, a continuous expression for local voxel porosity was calculated as an average weighted by the local phase membership probabilities.

A reactor network model is then used to combine the spatial distribution of the voxel porosity with the calcite dissolution rate in an imposed flow field. Each voxel is modelled as a mixed flow reactor (MFR) and is connected to the 6 neighboring voxels by plug flow reactors (PFR). The pressure drop occurs in PFR and follows Darcy's law

$$q = Rz \cdot l_n \varphi^2 \cdot \Delta p, \qquad (3)$$

where $Rz$ is a dimensionless number characterizing the dependence of the PFR's permeability on the neighboring voxels.

$$Rz = \left( \frac{P_{ref} L_{ref}}{Q_{ref}} \right) \cdot \left( \frac{a}{\mu} \right). \qquad (4)$$

The parabolic term stems from the hypothetical linear dependence of both permeability and tortuosity on voxel porosity when the characteristic length is 25 nm. The geometric mean of the porosities connected by the PFR is used. Assuming an incompressible fluid and continuity, the resulting system of 159.56 million equations was first solved using the stabilized biconjugate gradient method. The performance of the PFR, which relates the reactant concentration with the mixing state in the reactor, is given by

$$-C_0 + \frac{k'}{k' + k^I \left(1 - e^{Da \cdot k'}\right)} \cdot C = 0, \qquad (5)$$

where $Da$ the Damköhler number is defined as

$$Da = Hn \cdot (1 - \eta) \cdot \left( \frac{l_n^2}{q} \right) \cdot |\Delta \varphi|. \qquad (6)$$

The dimensionless number $Hn$ represents the sensitivity of calcite dissolution rate to aqueous calcium concentration given the maximum chemical affinity (reacting far from equilibrium)

$$Hn = \left( \frac{L_{ref}^2}{Q_{ref}} \right) \cdot k_A^0. \qquad (7)$$

$\eta$ represents the portion of the voxel volume modelled as an MFR. We use $\eta$ to replace Péclet number in the grey scale model because of the difficulties of obtaining the apparent diffusivities in voxels with varying porosity. $l_n^2 \cdot |\Delta \varphi|$ represents the surface area in the PFR, calculated as the spatial variation of material density across the voxel boundary. The performance equation for the MFR is

$$-\sum_i q_i C_{0,i} + \frac{1 - Da \cdot k'}{1 - Da \cdot (k^I + k')} \cdot qC = 0, \qquad (8)$$

where the Damköhler number is defined as

$$Da = Hn \cdot \eta \cdot \left(\frac{l_n^2}{q}\right) \cdot \sum_{i=1}^{6} |\Delta \varphi|_i . \qquad (9)$$

The summation represents the contribution of geometric surface area from the 6 neighboring voxels. $k^I$ is the dimensionless apparent first order rate constant defined as $k^I = k_{A0,app}^I / k_A^0$, where $k_{A0,app}^I = r_A / (C_{A,eq} - C_{A0})$. $k'$ is the dimensionless sensitivity of the calcite dissolution rate to the calcium concentration and is defined as $k' = k'_{A0} / k_A^0$. Both $k^I$ and $k'$ are devised, i.e., the 159.56 million equations for reactor performance can be solved iteratively using Eqs. 5 and 8. Using the inlet concentration from the previous iteration as the initial guess, the nonlinear rate law of calcite dissolution was applied repeatedly until the Frobenius norm of the difference between the consecutive concentration fields is less than $10^{-6}$. In the numerical simulations, only the transport of total aqueous calcium was tracked. The pH and the saturation index of calcite in each MFR and at each axial position of each PFR were computed by assuming calcite as the only source of calcium in a closed compartment. The speciation calculation was conducted with PhreeqC. The rate of calcite dissolution is then computed according to the rate law from Pokrovsky et al (*72, 73*). After the chemical conversion in each reactor (PFR and MFR) is obtained, the voxel porosity was updated based on mass balancing:

$$d\varphi = Ds \cdot l_n^{-3} \cdot \sum_{i=1}^{7} q_i (C_{0,i} - C_i), \qquad (10)$$

where the dimensionless number $Ds$ demonstrates the equivalence between the acid capacity and the time step of choice

$$Ds = \left(\frac{Q_{ref}}{L_{ref}^3}\right) \cdot \left(\frac{M}{\rho}\right) \cdot (C_{A,eq} - C_{A,inj}) \cdot dt . \qquad (11)$$

An auto-adaptive algorithm was used to choose the time step $dt$ so the overall porosity change of the simulation domain is capped at 0.01.

$$\Delta \varphi = \iiint_V \frac{1}{Hn \cdot l_n^2} \cdot q \cdot \frac{C_0 - C}{C_0} \cdot d\mathbf{r} \leq 0.01 \qquad (12)$$

For the optimal numerical stability with the given chalk microstructure the referencing quantities in the simulation were chosen heuristically as

$$L_{ref} = \frac{(C_{A,eq} - C_{A,inj}) dt}{Ds} \cdot \left(\frac{M}{\rho}\right) \cdot \left(\frac{k_A^0}{Hn}\right) \qquad (13)$$

$$P_{ref} = Rz \cdot \left(\frac{\mu}{a}\right) \cdot \frac{(C_{A,eq} - C_{A,inj}) dt}{Ds} \cdot \left(\frac{M}{\rho}\right) \cdot \left(\frac{k_A^0}{Hn}\right)^2 \qquad (14)$$

$$Q_{ref} = \left[\frac{(C_{A,eq} - C_{A,inj})dt}{Ds} \cdot \left(\frac{M}{\rho}\right)\right]^2 \cdot \left(\frac{k_A^0}{Hn}\right)^3 \qquad (15)$$

Table 1 lists the dimension and value of quantities used in the greyscale model.

**Acknowledgments:**

**General**: We thank F. Saxlid for help with the design and manufacturing of the percolation cell, J. U. Hammel for help with data collection at beamline P05 of Petra III at DESY (Deutsches Elektronen-Synchrotron, Germany), P. Alessandra for help with data collection at the TOMCAT beamline at SLS (Swiss Light Source), PSI (Paul Scherrer Institut, Switzerland), Y. Zheng for help with data collection with the XRadia Versa 410 system at DTU (Technical University of Denmark, Denmark) and H. Suhonen at the ID22 beamline at ESRF (European Synchrotron Research Facility, France) for technical support. Beamtime was provided by C. Gundlach and by SPring-8, DESY, PSI and ESRF through peer reviewed proposals. We thank the Japan Synchrotron Radiation Research Institute for the allotment of beam time on beamline BL20XU of SPring-8 (Proposal 2015A1147).

**Funding:** Funding for this project was provided by European Union's Horizon 2020 research and innovation programme under the Marie Sklodowska-Curie grant agreement No 653241, Innovation Fund Denmark through the CINEMA project, as well as the Innovation Fund Denmark and Maersk Oil and Gas A/S through the $P^3$ project. Travel support for synchrotron experiments was received from the Danish Agency for Science, Technology and Innovation via Danscatt.

**Author contributions:** YY designed the research and built the greyscale model. YY and SH conducted the percolation experiments. YY, HOS, KND and KU collected the tomography data. SB processed the images and converted the signals into a porosity distribution. SLSS and HOS supervised the research. All authors discussed the results and commented on the manuscript.

**Competing interests:** The authors claim no competing interests.

**Data and materials availability:** All data needed to evaluate the conclusions in the paper are presented. Additional data related to this paper may be requested from the authors by emailing yiyang@nano.ku.dk.

**Figures and Tables**

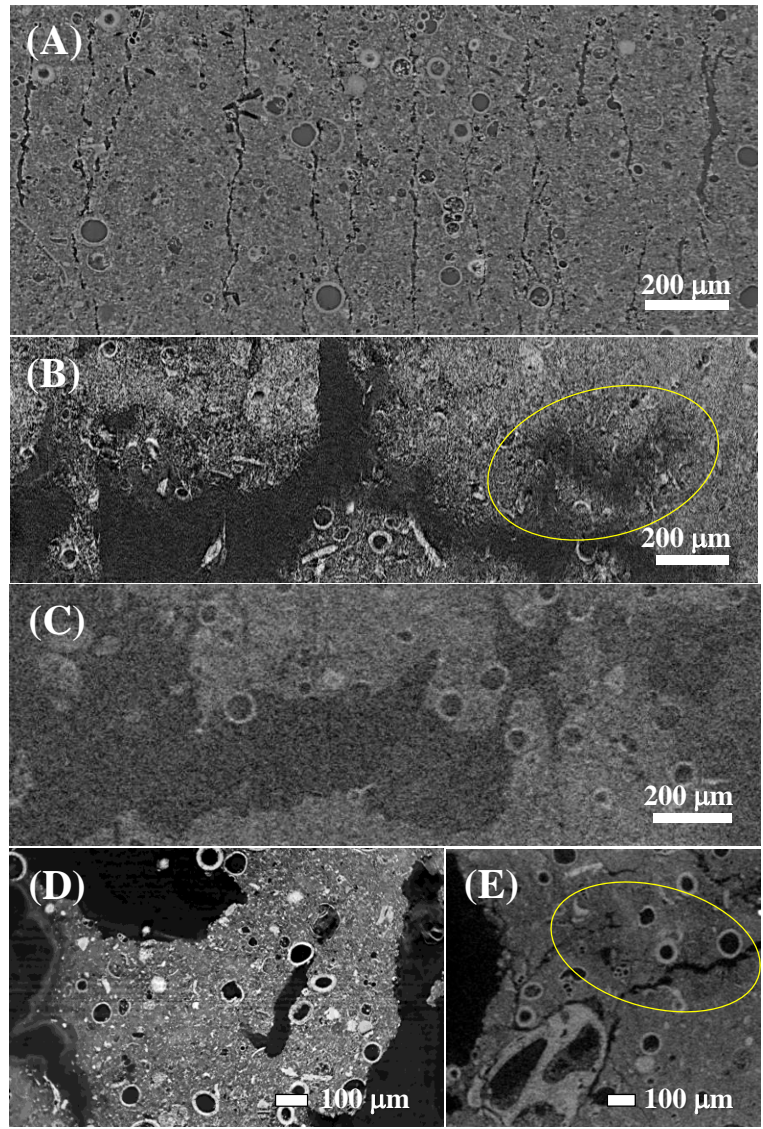

**Fig. 1. Morphology of wormholes in chalk.** (**A**) A representative texture of Aalborg chalk without wormholes on the micrometer scale and (**B**) – (**E**) wormholes formed during dissolutive percolation (flow direction: left to right). The same percolation condition (0.01 ml/min, 1 bar $CO_2$ dissolved in MilliQ water) can lead to different shapes of flow paths in different samples. (**B**) A growing wormhole captured by *in situ* X-ray tomography. The blurring edge and the advancing front of the pores (yellow circle) indicate dissolution during a tomographic scan. (**C**) A wormhole microstructure after fluid breakthrough. (**D**) A percolated sample (flow path beyond the viewing plane) with intact grain texture in the solid residual. (**E**) The presence of a large fossil near the fluid entrance diverts the growth of a wormhole towards paths with less resistance. Note the darker regions near the fractures (yellow circle) indicating partially removed materials. These features can be completely eliminated by image segmentation.

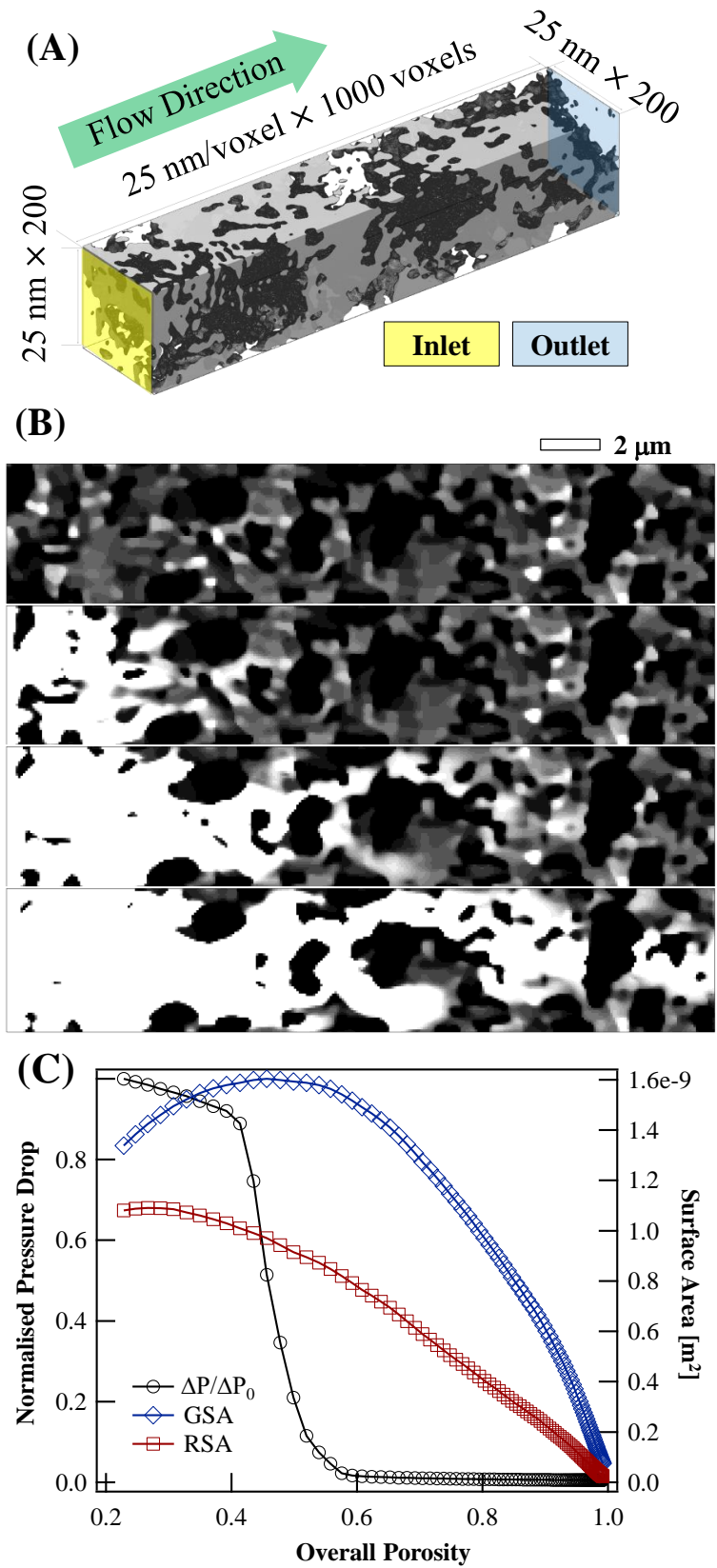

**Fig. 2. Simulated wormhole growth based on a 3D image of chalk determined by nanoCT.** (**A**) The simulation domain consists of 40 million voxels representing a 5 x 5 x 25 μm³ chalk cuboid. The isosurface is drawn at the 20% porosity. (**B**) Cross sections (5 x

25 µm$^2$) of the microstructure at 4 selected time steps where the average porosities of the volume were 0.2, 0.3, 0.4 and 0.5 (from top to bottom). The grey values of pixels indicate regional porosities and are scaled between 1 (voidspace, white) and 0 (pure solid, black). (**C**) Evolution of pressure drop across the sample in the flow direction (normalized to its initial value, $\Delta P/\Delta P_0$, left axis), geometric surface area (GSA, right axis) and reactive surface area (RSA, right axis).

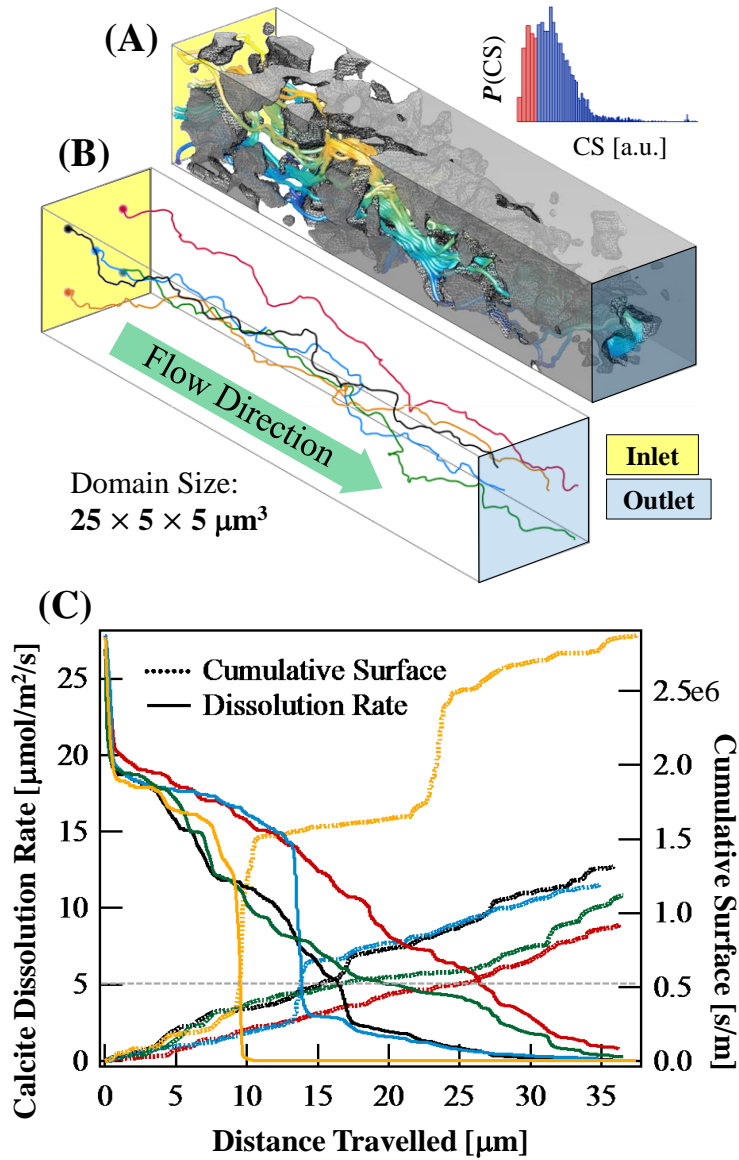

**Fig. 3. Impact of cumulative surface on pore growth in a numerical simulation.** (**A**) The microstructure at the cusp of breakthrough (porosity = 0.44, wormhole reaches the outlet) imposed on this are the streamlines that have been calculated based on the initial flow field (porosity = 0.20, intact microstructure). The isosurface is drawn at porosity = 1. The inset shows the shape of the cumulative surface distribution calculated based on 40,000 streamlines originating from the fluid inlet. The y-axis, P, is the normalized density function of CS. Only the streamlines within the lowest 10% of CS (red bars) are drawn. (**B**) Five streamlines originated from the fluid inlet. (**C**) The rate of calcite dissolution and the cumulative surface as functions of the distance fluid elements travel along the streamlines. The colors correspond to the streamlines in (**B**). The streamlines have similar

lengths and thus tortuosity, but differ greatly in their cumulative surface and the rate profiles. The grey dashed line cuts the curves at the positions where the dissolution rate in the streamlines drops to 5 µmol·m$^{-2}$·s$^{-1}$.

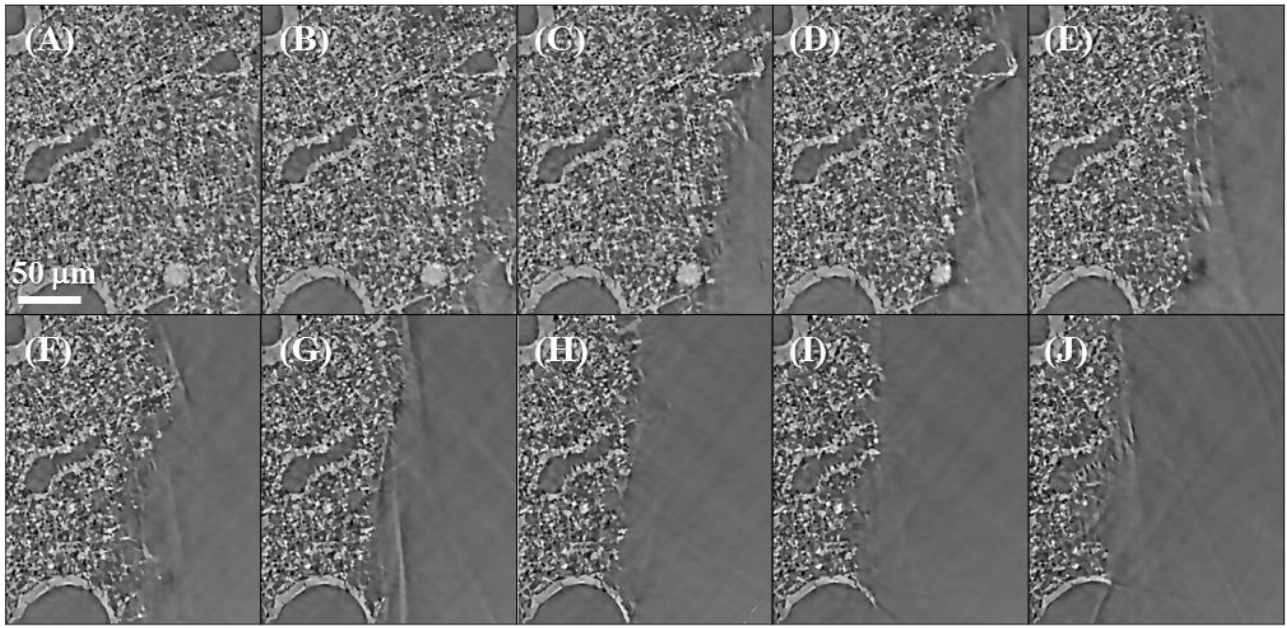

**Fig. 4. Generation and expansion of a wormhole in chalk.** (A) – (J) represent snapshots from a time series of reconstructed images recorded with *in situ* X-ray tomography over 21 hours. Shown are cross sections of a 200×245×300 µm$^3$ subvolume of a chalk sample undergoing dissolutive percolation (Perspective view in Fig. 5). The flow direction is normal to the plane of display pointing away from the reader. A wormhole originated from the lower right of the region of interest (ROI) and expanded left upwards while advancing along the flow direction. Solid material along flow paths with small cumulative surface (near the pore wall) is removed preferentially whereas the grain texture on the left of the observed region of interest is preserved.

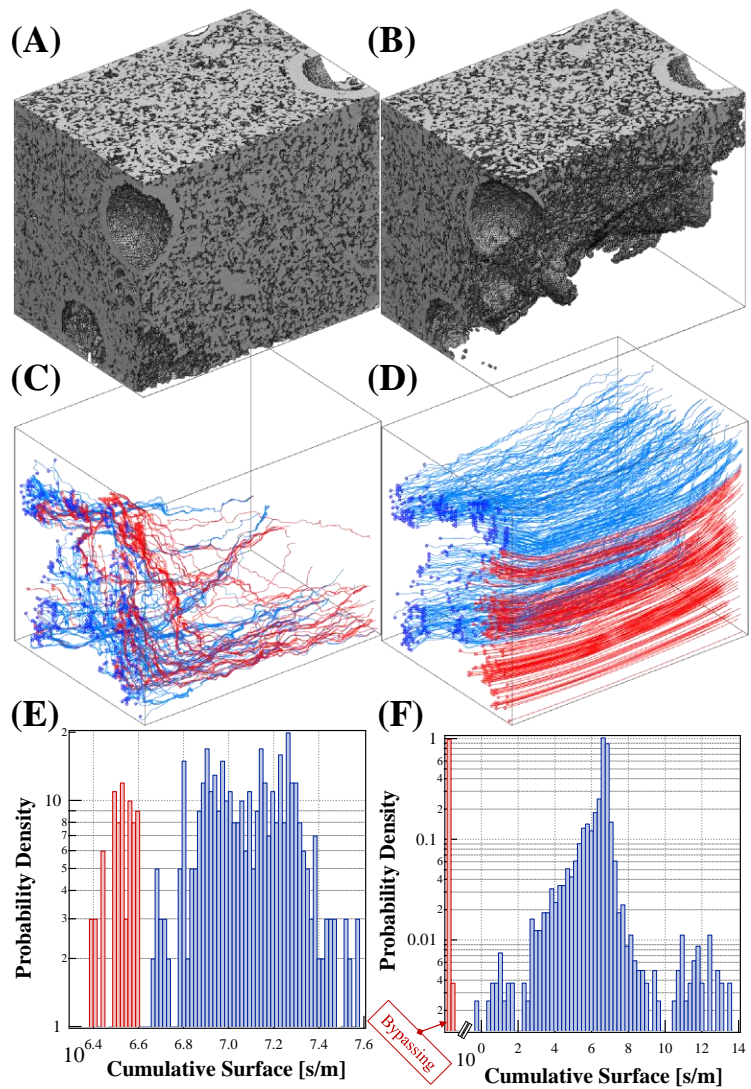

**Fig. 5. Impact of cumulative surface on experimentally observed flow channel development.** (**A**) and (**B**) The same region of microstructure as in Fig. 4 (200×245×300 μm$^3$) before and after 21 hours of percolation with carbonated MilliQ water (equilibrated with 1 bar $CO_2$). (**C**) Streamlines shown originate from the fluid inlet with the lowest 10% of cumulative surface, among which the lowest 2% are colored red. The streamlines are computed based on the flow field within the initial microstructure. (**D**) Streamlines originating from the same points depicted in (**C**) based on the flow field in the final microstructure. (**E**) and (**F**) Distributions of cumulative surface before and after the percolation. The long tails of both distributions consisting 10% of the total probability towards greater cumulative surface are not shown. The probability density functions are normalized to the logarithm of cumulative surface value so the integral area of the probability density function is 1. The red bars relate to the red streamlines in figures (C) and (D). In the x-axis in (F) there is a gap below 1 because the bypassing fluid elements have cumulative surface values close or equal to zero.

**Table 1. List of symbols for the greyscale model.** Numbers are given when a single value is used for all reactors. The dimensionless groups are calculated based on a timestep of 60 seconds.

| Symbol | Explanation | Dimension | Value |
|---|---|---|---|
| $\|\Delta\varphi\|$ | Porosity difference between two voxels connected by a PFR | | |
| $\|\Delta\varphi\|_i$ | Porosity difference between the voxel in question and its $i$th neighbor | | |
| $a$ | A lumping coefficient reflecting the dependence of permeability and tortuosity on voxel porosity | m² | $10^{-13}$ |
| $C$ | Dimensionless Ca concentration at reactor outlet | | |
| $C_0$ | Dimensionless Ca concentration at reactor inlet | | |
| $C_{0,i}$ | Dimensionless Ca concentration of the $i$th inlet of a MFR | | |
| $C_{A,eq}$ | Equilibrium concentration of Ca | mol/m³ | 0.0084 |
| $C_{A,inj}$ | Injection concentration of Ca | mol/m³ | 0 |
| $C_{A0}$ | Ca concentration at reactor inlet | mol/m³ | |
| $Da$ | Damköhler number (Eqs. 6 & 9) | | |
| $Ds$ | Denis number (Eq. 11) | | 0.0074 |
| $dt$ | Timestep | s | 60 |
| $\eta$ | Voxel level mixing factor varying from 1 (complete) to 0 (no) mixing. | | 0.5 |
| $Hn$ | Henning number (Eq. 7) | | 332.5485 |
| $\varphi$ | Porosity | | |
| $k'$ | Dimensionless first order rate constant | | |
| $k_A^0$ | First order rate constant at the injection concentration of Ca | m/s | 0.0033 |
| $k^I$ | Dimensionless linearized rate constant | | |
| $k^I_{A0,app}$ | Linearized rate constant | m/s | |
| $l$ | Voxel dimension | nm | 25 |
| $l_n$ | Normalized voxel dimension | | 1 |
| $L_{ref}$ | Characteristic length | m | $2.5\times10^{-8}$ |
| $\mu$ | Fluid viscosity | Pa·s | $8.9\times10^{-4}$ |
| $M$ | Molar density | kg/mol | 0.1 |
| $p$ | Dimensionless pressure | | |
| $P_{ref}$ | Reference pressure | Pa | 1 |
| $q$ | Dimensionless volumetric flowrate | | |
| $q_i$ | Dimensionless volumetric flow rate from the $i$th inlet of an MFR | | |
| $Q_{ref}$ | Reference volumetric flowrate (per inlet voxel) | m/s | $6.25\times10^{-21}$ |
| $\rho$ | Density | kg/m³ | 2710 |
| $r_A$ | Calcite dissolution rate | mol/m²/s | |
| $Rz$ | Reza number (Eq. 4) | | 449.4382 |